\documentclass[preprint,
,nofootinbib
]{revtex4-1}
\usepackage{epstopdf}
\usepackage{graphicx}
\usepackage{subfigure}
\usepackage{color}
\usepackage{amsmath}
\usepackage{epsfig,graphicx,bm}
\usepackage{epstopdf, epsf}
\usepackage{dcolumn}
\usepackage{hyperref}
\everymath{\displaystyle}
\begin{document}
\title{Notes on diffeomorphisms
symmetry of $f(R)$ gravity in the cosmological context}
\author{Amir Ghalee}
\affiliation{{Department of Physics, Tafresh  University,
P. O. Box 39518-79611, Tafresh, Iran}}
\begin{abstract}
We study the metric perturbations in the context of restricted $f(R)$ gravity, in which a parameter for deviation from the full diffeomorphisms of space-time is introduced. We demonstrate that one can choose the parameter to remove the induced anisotropic stress, which is present in the usual $f(R)$ gravity. Moreover, to prevent instability for the vector and tensor metric perturbations, some constraints on the restricted $f(R)$ gravity are obtained. 
\end{abstract}
\maketitle
\section{INTRODUCTION}
Plank 2015 data show that the so-called $R^{2}$ inflation model for the early Universe is consistent with the observations \citep{plank-data}. Also, $f(R/M^{2})$-gravity, which is described by
\begin{equation}\label{i-0}
S_{f}=\int d^{4}x\sqrt{-g}\left[\frac{M_{P}^{2}R}{2}+\lambda M_{P}^{4}f(\frac{R}{M_{P}^{2}})\right]+S_{Matter},
\end{equation}
where $M_{P}^{2}$ is the reduced Planck
mass and $\lambda$ is a dimensionless constant, have been proposed for the late-time cosmology to avoid the cosmological constant problem \cite{modi}. Such observations and theoretical speculations provide motivations to investigate the modified gravity theories in the context of cosmology.\\
Study of the metric perturbations provide a tool to characterize the modified gravity theories. For example, if we consider $f(R/M^{2})$-gravity in the perturbed Friedmann-Robertson-Walker(FRW) Universe and using the Newtonian gauge, which is defined by
\begin{equation}
ds^{2}=-(1+2\Phi(t,x))dt^{2}+a(t)^{2}(1+2\Psi(t,x))dx^{i}dx^{j}\delta_{ij},
\end{equation}
we lead to $\Phi(t,x)\neq\Psi(t,x) $, i.e. the induced anisotropic stress arises. This effect can be used to distinguish the modified gravity theories from the observations \cite{silk}.\\
One way to generalize \ref{i-0} is to use other curvature invariant quantities in four-dimensions, such as $R, R_{\mu\nu}R^{\mu\nu},...$ \cite{modi1}.
On other hand, in the cosmological context, the cosmic microwave background shows that the FRW metric is the preferred coordinate system of the Universe. Thus, to describe dynamics of the Universe, one can consider theories which are invariant just under the spacial diffeomorphisms as have been done in \cite{chen} and references therein. So, one can also apply this idea to $f(R/M^{2})$-gravity and restrict the symmetry of the action to just under the spacial diffeomorphisms.\\
In this paper we study the relation between the induced anisotropic stress and the full diffeomorphisms symmetry of $f(R/M^{2})$-gravity. as we will see, a systematic way to obtain this relation is to construct restricted $f(R/M^{2})$ from the usual $f(R/M^{2})$. The main motivation to report this work comes from one of the interesting results of this attempt that is given by Eq. \ref{im5}. The result shows that how the induced anisotropic stress is related to the full diffeomorphisms symmetry. Note that, as we argued, in the cosmological context the full diffeomorphism symmetry, is broken by using FRW metric as the preferred coordinate. For this purpose we impose the following transformation in \ref{i-0}, which break the time diffeomorphisms of \ref{i-0} but save the symmetry under the spacial diffeomorphisms,
\begin{equation}\label{i-2}
R\rightarrow R_{\Upsilon}\equiv R+(\Upsilon-1)\Xi,
\end{equation}
where $\Upsilon$ is a parameter and $\Xi$ is a four-divergence term, which is appear in the decomposition of the  the Ricci scalar in four dimensions as \cite{wald}
\begin{equation}\label{i-1}
R={}^3R+(K^{ij}K_{ij}-K^{2})+\Xi,
\end{equation}
where $^3R$ is the three-dimensional Ricci scalar which is obtained from a three-dimensional metric $h_{ij}$. Also $K=h^{ij}K_{ij}$, where $K^{ij}$ is the extrinsic curvature,
which is defined as
\begin{equation}
K_{ij}=\frac{1}{2N}(\dot{h}_{ij}-\nabla_{i}N_{j}-\nabla_{j}N_{i}),
\end{equation}
where $N$ and $N^{i}$ are the laps and shift respectively.\\
To state $\Xi$ in terms of the laps and the shift, we use the general formula in Ref. \citep{wald}, in which one finds
\begin{equation}\label{refad}
\Xi=2\nabla_{\mu}(n^{\mu}K)-\frac{2}{N}\nabla_{i}\nabla^{i}N,
\end{equation}
where $ n_{\mu} $ can be represented as
\begin{equation}
n_{\mu}=(-N,0,0,0).
\end{equation}
Note that although the last term in \ref{i-1} is the four-divergence term and for the Einstein-Hilbert action
dose not affect field equations in 3+1 formalism, but without it
$R$ is just invariant under the spacial diffeomorphisms.\\
Let us urge the reader that our purpose is to find the relation between the induced anisotropic stress and the diffeomorphism symmetry in $f(R/M^{2})$-gravity. It is important point because If one wants to regard this attempt as a new model, one must quest for the Hamiltonian consistency of the model. To see this approach see Ref. \cite{chaic-gh-kl}. Also, to see a different motivations to apply the above transformation in the context of Horava-Lifshitz gravity, see Ref. \cite{chai}.\\
The organization of this paper is as follows: in Sec. 2 we briefly review background cosmology of the restricted $f(R/M^{2})$-gravity. Sec. 3 is devoted to study the dynamics of the model with perturbed metric.
\section{Background equations}
In this section, we obtain general equations for the restricted $f(R/M^{2})$-gravity for the unperturbed FRW metric background. Note that in all relations and results of this work, if we take $\Upsilon=1$ the corresponding relations for the usual $f(R/M^{2})$-gravity must be obtained.\\
By applying the transformation that is shown in \ref{i-2},
the restricted version of any $f(R/M^{2})$-gravity is obtained from \ref{i-0} as
\begin{equation}\label{ac1}
S_{res}=\int d^{4}x\sqrt{-g}\left[M_{P}^{2}\frac{R}{2}+\lambda M_{P}^{4}f\big(\frac{R_{\Upsilon}}{M_{P}^{2}})\right]+S_{Matter}.
\end{equation}
We take the flat FRW metric as
\begin{equation}\label{1-1}
ds^{2}=-N(t)^{2}dt^{2}+a(t)^{2}dx^{i}dx^{j}\delta_{ij},
\end{equation}
where $N=N(t)$ is the laps and $a=a(t)$ is the scale factor that from which the Hubble parameter is defined as $H\equiv\dot{a}/a$.\\
The energy-momentum tensor of the prefect fluid, $T_{\mu\nu}$, is obtained by
\begin{equation}\label{i-3}
T_{\mu\nu}=-\frac{2}{\sqrt{-g}}\frac{\delta S_{Matter}}{\delta g^{\mu\nu}}.
\end{equation}
The matter is the prefect fluid which is minimally coupled to the metric. Without any non-minimal interaction between gravity and the matter, we breaks   the general diffeomorphism just in the gravitational sector of the action. In Ref. \citep{addref2} it has been shown that if one just breaks the general diffeomorphism for the gravitational sector, the usual conservation of energy, $\nabla^{\mu}T_{\mu\nu}=0 $, is hold. So, by using this equation for the FRW metric, it follows that
\begin{equation}\label{i-33}
\dot{\rho}+3H(\rho+p)=0,
\end{equation}
where $\rho\equiv-T_{0}^{0}$ and $p\delta_{i}^{j}\equiv T_{i}^{j}$ are the density of energy and momentum of the fluid respectively.\\
Using the above metric and the general relations in the ADM formalism, which are discussed in Ref. \citep{wald}, one can show that
\begin{equation}\label{i-4}
R_{\Upsilon}=R+(\Upsilon-1)\Xi=A+\Upsilon\Xi,
\end{equation}
where
\begin{equation}\label{i-3}
\Xi=-6\frac{H\dot{N}}{N^{3}}+6\frac{\dot{H}}{N}+18 \frac{H^{2}}{N^{2}}\hspace{.08cm},\quad A\equiv-6\frac{H^{2}}{N^{2}}.
\end{equation}
It is easy to check that, by integration by parts, we have
\begin{equation}
\int d^{4}x a^{3}N\Xi=0,
\end{equation}
which, as we argued, shows that $\Xi$ is the four-divergence term.\\
From Eqs. \ref{ac1} and \ref{i-3}, it follows that
\begin{equation}
S_{res}=\int d^{4}x\sqrt{-g}\left[M_{P}^{2}\frac{R}{2}+\lambda M_{P}^{4}f\big(\frac{A+\Upsilon \Xi}{M_P^{2}}\big)\right]+S_{Matter}.
\end{equation}
Using Eqs. \ref{i-4} and \ref{i-3} and then varying the action with respect to the laps, with some integration by parts and then setting $N=1$, we have
\begin{equation}\label{gen11}
3H^{2}+\lambda M_{P}^{2}f+\lambda (6H^{2}-R_{\Upsilon}) F+6\lambda\Upsilon H \dot{F}=\frac{\rho}{M_{P}^{2}},
\end{equation}
where
\begin{equation}\label{fdef}
R_{\Upsilon}= -6H^{2}+\Upsilon (6\dot{H}+18H^{2}),\hspace{0.15cm} f\equiv f\big(\frac{R_{\Upsilon}}{M_{P}^{2}}\big)\hspace{0.15cm},F\equiv M_{P}^{2}\frac{df}{dR_{\Upsilon}}\equiv f'.
\end{equation}
Note that we have defined prime as derivative with respect to the argument of $f$. So, $F,f'',...$ are dimensionless quantities.\\
Also, an useful equation is obtained by taking the time derivative of Eq. \ref{gen11} and then using Eq. \ref{i-3} as
\begin{equation}\label{bae}
\dot{H}(1+2\lambda F)+\lambda H (2-3\Upsilon)\dot{F}+\Upsilon \lambda\ddot{F}=-\frac{(\rho+p)}{2M_{P}^{2}},
\end{equation}
Present of the term which is proportional to $\ddot{F}$, shows that in general we will confront with the fourth order differential equations for the scale factor in this model. Similar to the usual $f(R/M_P^{2})$-gravity if we neglect the matter or impose some additional symmetry, like the de Sitter space-time, one can reduce the order of the equations.
\section{\label{sec:level1}  Cosmological perturbations}
In this section, to obtain equations for the perturbed FRW space-time, we will use the Arnowitt-Deser-Misner(ADM) formalism. for this goal, the general formula that discussed in Ref. \citep{wald} will be used.If one takes $\Upsilon=1$ in all relations and results, they must approach the corresponding results for the usual $f(R/M_{P}^2)$ gravity. Our notations in this work are similar to those that have been used for the usual $f(R/M_{P}^2)$ gravity by De Felice et al. \cite{defs}.\\
In the ADM formalism, a metric can be decomposed as \citep{wald}
\begin{equation}\label{c1}
ds^{2}=-N^{2}dt^{2}+h_{ij}(dx^{i}+N^{i}dt)(dx^{j}+N^{j}dt).
\end{equation}
From \ref{i-2} and \ref{refad}, we have
\begin{equation}\label{per1}
R_{\Upsilon}={}^{3}R+K^{ij}K_{ij}-K^{2}+2\Upsilon\nabla_{\mu}(n^{\mu}K)-\frac{2\Upsilon}{N}\nabla_{i}\nabla^{i}N.
\end{equation}
Note that the covariant derivative in the last term of Eq. \ref{per1} is taken with respect to $h_{ij}$ \citep{wald}.\\
As for the prefect fluid, we will use the following parameterization for the perturbed energy-momentum tensor
\begin{equation}\label{fl1}
\delta T_{0}^{0}=-\delta \rho,\hspace{.1cm}\delta T_{0}^{i}=-(\rho+p)\partial_{i}v,\hspace{.1cm}\delta T_{i}^{j}=\delta p\delta_{i}^{j},
\end{equation}
where $v$ is the potential for the spatial velocity of the fluid.\\
we will write our equations in the Newtonian gauge which is defined by $N^{i}=0$. Also, the equations will be written in the Fourier space for which the Fourier components of a general perturbation $U(t,\textbf{x})$ is given by
\begin{equation}
U=\int d^{3}xU(t,\textbf{x})e^{-i\textbf{k.x}}.
\end{equation}
Furthermore, $F$ and $\dot{F}$ can be decomposed into homogeneous and perturbed parts as
\begin{equation}
F=\bar{F}+\delta F,\hspace{.15cm} \dot{F}=\dot{\bar{F}}+\dot{\delta F},
\end{equation}
where $\hspace{0.1cm}\bar{ }\hspace{0.1cm}$ over any quantity shows the unperturbed part of the quantity.
\subsection{\label{sec:level1}  The scalar metric perturbations}
It is convenient that parameterize the the scalar metric perturbations in the Newtonian gauge as
\begin{equation}
N=e^{\Phi(t,\textbf{x})},\hspace{.15cm} h_{ij}=a^{2}e^{-2\Psi(t,\textbf{x})}\delta_{ij}.
\end{equation}
From \ref{per1} and the above definitions, it follows that
\begin{equation}
\begin{split}
R_{\Upsilon}|_{scalar}&={}^{3}R+6(3\Upsilon-1)e^{-2\Phi(t,\textbf{x})}(H-\dot{\Psi}(t,\textbf{x}))^{2}\\
&+6\Upsilon e^{-2\Phi(t,\textbf{x})}(\dot{H}-\ddot{\Psi}(t,\textbf{x}))\\
&-6\Upsilon\dot{\Phi}(t,\textbf{x})e^{-2\Phi(t,\textbf{x})}
(H-\dot{\Psi}(t,\textbf{x}))\\
&-2\Upsilon\frac{e^{2\Psi(t,\textbf{x})}}{a^{2}}\partial^{2}\Phi(t,\textbf{x})
+2\Upsilon\frac{e^{2\Psi(t,\textbf{x})}}{a^{2}}\partial_{i}\Phi(t,\textbf{x})\partial_{i}\Psi(t,\textbf{x})\\
&-2\Upsilon\frac{e^{2\Psi(t,\textbf{x})}}{a^{2}}\partial_{i}\Phi(t,\textbf{x})\partial_{i}\Phi(t,\textbf{x}).
\end{split}
\end{equation}
If we take $\Phi(t,\textbf{x})=\Psi(t,\textbf{x})=0$ in the above relation, it follows that ${}^{3}R$ is vanished and the value of $R_{\Upsilon}$ is reduced to Eq. \ref{fdef}.\\
it is easy to show the linearized part of the above relation, $\delta R_{\Upsilon}|_{scalar}$, takes the following form (in terms of the Fourier components)
\begin{equation}\label{ldr}
\begin{split}
\delta R_{\Upsilon}|_{scalar}=&\frac{-4k^2}{a^{2}}\Psi+\frac{2\Upsilon k^{2}}{a^{2}}\Phi-12(3\Upsilon-1)H^{2}\Phi\\
&-12\Upsilon \dot{H}\Phi-6\Upsilon H\dot{\Phi}-6\Upsilon\ddot{\Psi}-12(3\Upsilon-1)H
\dot{\Psi}.
\end{split}
\end{equation}
Since the fluid is minimally coupled to the gravity, from $\nabla^{\mu}T_{\mu\nu}=0$ (which at the background level results in Eq. \ref{i-33}), we can obtain two equations. They are the same as the corresponding equations for the usual $f(R/M^{2})$ gravity as
\begin{equation}\label{im1}
\dot{\delta\rho}+3H(\delta\rho+\delta p)=\frac{k^2}{a^2}\delta q +3(\rho+p)\dot{\Psi},
\end{equation}
and
\begin{equation}\label{im2}
\dot{\delta q}+3H\delta q+\delta p+(\rho+p){\Phi}=0.
\end{equation}
where $\delta q\equiv-(\rho+p)v$.\\
For other equations, we must obtain the second order action by inserting \ref{c1} into the action. As we pointed out, we have used the Newtonian gauge. So, to
variate the action with respect to the shift, it is sufficient to consider the terms which are proportional to $N_{i}\Psi$ and $N_{i}\Phi$.  For example if we define $\delta_{N_{i}}$ as variation with respect to the shift, for $S_{Matter}$ we have
\begin{equation}
\delta_{N_{i}} S_{Matter}=\frac{-1}{2}\int d^{4}x\sqrt{-g}T^{\mu\nu}\delta_{N_{i}} g_{\mu\nu}=\frac{-1}{2}\int d^{4}x a^{3}T^{0i}\delta N_{i}+\mathcal{O}(N_i^{2}).
\end{equation}
Using this point, and also use the following relations
\begin{equation}
n^{\mu}=\big(e^{-\Phi(t,\textbf{x})},-N^{i}e^{-\Phi(t,\textbf{x})}\big),\hspace{.1cm} \Gamma^i_{ij}=-3\partial_{j}\Psi(t,\textbf{x}),
\end{equation}
after some integration by parts, we lead to
\begin{equation}
\begin{split}
S=&-M_{P}^{2}\int d^{4}x 2a[H\Phi(t,\textbf{x})+\dot{\Psi}(t,\textbf{x})]\partial_{i}N_{i}(1+2\lambda\bar{F})\\
&-4\lambda M_{P}^{2}\int d^{4}x a H N_{i}\partial_{i}\delta F\\
&-2\lambda\Upsilon M_{P}^{2}\int d^{4}x a[N_{i}\partial_{i}\dot{\delta F}-\partial_{i}\Phi(t,\textbf{x}) N_{i}\dot{\bar{F}}-3H N_{i}\partial_{i}\delta F]+ S_{Matter}
\end{split}
\end{equation}
Thus, variation with respect to the shift, and then using the Fourier components of the perturbations, yields
\begin{equation}\label{im3}
(H\Phi+\dot{\Psi})(1+2\lambda\bar{F})=\lambda\big[\Upsilon\dot{\delta F}-\Upsilon
\Phi\dot{\bar{F}}+(2-3\Upsilon)H\delta F\big]
-\frac{1}{2M_{P}^{2}}\delta q.
\end{equation}
Now to variate the action with respect to $\Phi(t,\textbf{x})$, we can set $N_{i}=0$ in the action and then expand the action to the second order in $\Phi(t,\textbf{x})$ and $\Psi(t,\textbf{x})$. This procedure, after some integration be parts and using Eq. \ref{gen11}, leads to
\begin{equation}
\begin{split}
\delta_{\Phi}S_{res}=\int &d^{4}x a^{3} M_{P}^{2}\delta\Phi(t,\textbf{x})\Big[(-12\Phi(t,\textbf{x}) H^{2}+\frac{4}{a^{2}}\partial^{2}\Psi(t,\textbf{x})\\
&-12H\dot{\Psi}(t,\textbf{x}))(\frac{1}{2}+\lambda\bar{F})\\
&+\lambda\big((12-18\Upsilon)H^{2}-6\Upsilon\dot{H}-\frac{2\Upsilon}{a^{2}}\partial^{2}\big)\delta F\\
&-\lambda\dot{\bar{F}}\big(12\Upsilon H\Phi(t,\textbf{x})+6\Upsilon\dot{\Psi}(t,\textbf{x})\big) +
6\lambda\Upsilon H\dot{\delta F}\\
&-\delta_{\Phi}S_{Matter}\Big].
\end{split}
\end{equation}
Thus, variation withe respect to $\Phi(t,\textbf{x})$ and then using the Fourier space leads us to
\begin{equation}\label{im4}
\begin{split}
&\lambda\big [(6-9\Upsilon)H^{2}-3\Upsilon\dot{H}+\frac{\Upsilon}{a^{2}}k^{2}\big]\delta F
-3\Upsilon\lambda\dot{\bar{F}}(2H\Phi+\dot{\Psi})\\
&+
3\lambda\Upsilon H\dot{\delta F}-\frac{\delta\rho}{2M_{P}^{2}}
=(3\Phi H^{2}+\frac{k^{2}}{a^{2}}\Psi +3H\dot{\Psi})(1+2\lambda\bar{F})
\end{split}
\end{equation}
For reasons that will become clear, we will obtain $\delta p$ from two ways.
For the first way, Using Eqs. \ref{im1}, \ref{im4} and also Eq. \ref{bae} to eliminate $(\rho+p)$ in these formula, leads us to
\begin{equation}\label{p1}
\begin{split}
\frac{\delta p}{2M_P^{2}}=&\lambda(2\dot{\Psi}+4H\Phi+\Upsilon\dot{\Phi})
\dot{\bar F}-2\lambda H\dot{\delta F}+2\lambda\Upsilon\Phi\ddot{\bar F}-\lambda\Upsilon\ddot{\delta F}\\
&+\lambda\big[(3\Upsilon-2)\dot{H}+(9\Upsilon-6)H^{2}+\frac{2k^{2}}{3a^{2}}(1-2\Upsilon)\big]\delta F\\
&+\big[\dot{\Phi}H+2\dot{H}\Phi+\ddot{\Psi}+3\Phi H^{2}+3H\dot{\Psi}\\
&+\frac{k^2}{3a^{2}}(\Psi-\Phi)\big](1+2\lambda \bar{F}).
\end{split}
\end{equation}
Also, to obtain the above equation, we have used the fact that $ \dot{R}_\Upsilon\delta F=\delta R_\Upsilon\dot{\bar{F}} $.\\
The other relation for $\delta p$ is obtained by using Eqs. \ref{im2}, \ref{im3}. Again, after using Eq. \ref{bae} to eliminate $(\rho+p)$, we have
\begin{equation}\label{p2}
\begin{split}
\frac{\delta p}{2M_P^{2}}=&\lambda(2\dot{\Psi}+4H\Phi+\Upsilon\dot{\Phi})
\dot{\bar F}-2\lambda H\dot{\delta F}+2\lambda\Upsilon\Phi\ddot{\bar F}-\lambda\Upsilon\ddot{\delta F}\\
&+\lambda\big[(3\Upsilon-2)\dot{H}+(9\Upsilon-6)H^{2}\big]
\delta F\\
&+\big[\dot{\Phi}H+2\dot{H}\Phi+\ddot{\Psi}+3\Phi H^{2}+3H\dot{\Psi}\big](1+2\lambda \bar{F}).
\end{split}
\end{equation}
The right-hand side of Eqs. \ref{p1} and \ref{p2} are the same if
\begin{equation}\label{im5}
(\Psi-\Phi)(1+2\lambda \bar{F})=2\lambda(2\Upsilon-1)\delta F.
\end{equation}
As we pointed out if we take $\Upsilon=1$ in the above equations, they must approach to the corresponding relations for the usual $f(R/M_{P}^{2})$ gravity. Note that, if we consider the prefect fluid, we have four independent variables in the model and only four of the above equations are independent.\\
Eq. \ref{im5} reveals an advantage of our formalism. One
of the features in almost all modified gravity theories is existence of the induced anisotropic stress which shows itself by $\Phi\neq\Psi$. As is clear from Eq. \ref{im5} one can chose $\Upsilon=1/2$ to eliminate the anisotropic stress in our model.
\subsection{\label{sec:level1}  The tensor metric perturbations}
The tensor metric perturbations,$\gamma_{ij}$, are characterized by
\begin{equation}
ds^{2}=-dt^{2}+a^{2}[\delta_{ij}+\gamma_{ij}]dx^{i}dx^{j},
\end{equation}
where $\partial_{i}\gamma_{ij}=\gamma_i^i=0$. From the above definition and the traceless condition on $\gamma_{ij}$, it turns out that the terms in \ref{per1} which are proportional to $\Upsilon$ do not contribute to the tensor metric perturbations. So, study of this sector is very similar to the usual $f(R/M^{2})$ gravity. The second order action for this sector becomes
\begin{equation}\label{t1}
\delta S|_{tensor}=\frac{M_{P}^{2}}{8}\int d^{4}x[1+2\lambda F]\big[a
\gamma_{ij}\partial^{2}\gamma_{ij}+a^{3}\dot{\gamma}_{ij}^{2}
\big].
\end{equation}
Variate \ref{t1} with respect to $\gamma_{ij}$ and using
the following Fourier representation
\begin{equation}\label{1-26}
 \gamma_{ij}=\int\frac{d^{3}k}{(2\pi)^{3/2}}\sum_{s=\pm}\epsilon_{ij}^{s}(k)\gamma_{k}^{s}(t)e^{i\overrightarrow{k}.\overrightarrow{x}},
\end{equation}
 where $\epsilon_{ii}=k^{i}\epsilon_{ij}=0$ and $\epsilon_{ij}^s(k)\epsilon_{ij}^{{s'}}(k)=2\delta_{s'}$, leads to
  \begin{equation}\label{1-27}
\ddot{\gamma_{k}^{s}}+\dot{\gamma_{k}^{s}}\frac{d}{dt}\ln[a^{3}(1+2\lambda F)]+(\frac{k}{a})^{2}\gamma_{k}^{s}=0.
\end{equation}
The second term in Eq. \ref{1-27} must be positive to act as a dissipative forces. Otherwise, $ \gamma_{k}^{s} $  grows without bound and we will confront with instability in this sector. Also, similar to discussion for the usual $ f(R/M_p^{2}) $-gravity in Ref.\cite{defs}, to avoid the ghost instability, we must impose the following condition
\begin{equation}\label{s1}
1+2\lambda F>0.
\end{equation}
Thus, from the above discussion and Eq. \ref{s1}, we have
 \begin{equation}\label{s1}
\lambda F'>0.
\end{equation}
Again, the above conditions are similar to the corresponding relation in the usual $f(R/M_p^{2})$-gravity.
\subsection{\label{sec:level1}  The vector metric perturbations}
As the prefect fluid, we consider the following form for the perturbed stress-tensor in the vector sector
\begin{equation}
\delta T^{0}_{i}|_{vector}=\delta q_{i}^{V}.
\end{equation}
So, from $\nabla_{\mu}T^{\mu}_{\nu}=0$ it follows that
\begin{equation}
\dot{\delta q_{i}^{V}}+3H\delta q_{i}^{V}=0.
\end{equation}
As for the metric perturbations, the  favorite gauge in this sector is the so-called vector gauge which is defined by
\begin{equation}
ds^{2}=-dt^{2}+2aS_{i}dx^{i}dt+a^{2}\delta_{ij}dx^{i}dx^{j},
\end{equation}
where $\partial_{i}S_{i}=0$. Again, from the above definition and the condition on $S_{i}$, it turns out that the terms in \ref{per1} which are proportional to $\Upsilon$ do no any effect in this sector. Also, the second order action takes the following form
\begin{equation}\label{vas}
\delta S|_{vector}=-\frac{M_{p}^{2}}{2}\int d^{4}x aS_{i}\partial^{2}S_{i}(1+2\lambda F)-2\int d^{4}x a^{2}S_{i}\delta q_{i}^{V}.
\end{equation}
Thus, in the Fourier space the equation for $S_{i}$ is
\begin{equation}
M_{p}^{2}k^{2}(1+2\lambda F)\frac{S_{i}}{a}=2\delta q_{i}^{V}.
\end{equation}
Also, from Eq. \ref{vas}, to avoid the ghost instability in the vector metric perturbations, it is sufficient to take the condition which we have in Eq. \ref{s1}
\section{\label{sec:level1} Summary}
We have studied the consequences of the systematic way to break the time diffeomorphisms of the usual $f(R/M_{P}^2)$ gravity. By investigating the cosmological perturbation of the restricted $f(R/M_{P}^2)$ gravity, we have shown that one can relate this symmetry to the induced anisotropic stress that arises in the usual $f(R/M_{P}^2)$. We have shown that one can choose $\Upsilon$, which is defined in \ref{i-2}, to eliminate the induced anisotropic stress. So, even if one dose not interested in the phenomenological consequences of the model, this work provide a tool to gain inside the dynamics of the metric perturbations of the usual $f(R/M_{P}^2)$ the modified gravity. Also, we have obtained some constraints on the model by demanding absent of the ghost instability in the tensor and vector metric perturbations.\\

\end{document}